# Solar coronal magnetic fields derived using seismology techniques applied to omnipresent sunspot waves

**Authors:**
David B. Jess[1,2*], Veronika E. Reznikova[3], Robert S. I. Ryans[1], Damian J. Christian[2], Peter H. Keys[1,4], Mihalis Mathioudakis[1], Duncan H. Mackay[5], S. Krishna Prasad[1], Dipankar Banerjee[6], Samuel D. T. Grant[1], Sean Yau[1], Conor Diamond[1]

**Affiliations:**
[1]Astrophysics Research Centre, School of Mathematics and Physics, Queen's University Belfast, Belfast, BT7 1NN, UK.
[2]Department of Physics and Astronomy, California State University Northridge, Northridge, CA 91330, U.S.A.
[3]Center for Mathematical Plasma Astrophysics, Department of Mathematics, KU Leuven, Celestijnenlaan 200B bus 2400, B-3001 Heverlee, Belgium.
[4]Solar Physics and Space Plasma Research Centre (SP[2]RC), University of Sheffield, Hicks Building, Hounsfield Road, Sheffield, S3 7RH, UK.
[5]School of Mathematics and Statistics, University of St Andrews, St Andrews, KY16 9SS, UK.
[6]Indian Institute of Astrophysics, II Block, Koramangala, Bangalore 560 034, India.

*To whom correspondence should be addressed. Email: d.jess@qub.ac.uk

**Sunspots on the surface of the Sun are the observational signatures of intense manifestations of tightly packed magnetic field lines, with near-vertical field strengths exceeding 6,000 G in extreme cases[1]. It is well accepted that both the plasma density and the magnitude of the magnetic field strength decrease rapidly away from the solar surface, making high-cadence coronal measurements through traditional Zeeman and Hanle effects difficult since the observational signatures are fraught with low-amplitude signals that can become swamped with instrumental noise[2,3]. Magneto-hydrodynamic (MHD) techniques have previously been applied to coronal structures, with single and spatially isolated magnetic field strengths estimated as 9–55 G[4-7]. A drawback with previous MHD approaches is that they rely on particular wave modes alongside the detectability of harmonic overtones. Here we show, for the first time, how omnipresent magneto-acoustic waves, originating from within the underlying sunspot and propagating radially outwards, allow the spatial variation of the local coronal magnetic field to be mapped with high precision. We find coronal magnetic field strengths of 32 ± 5 G above the sunspot, which decrease rapidly to values of approximately 1 G over a lateral distance of 7000 km, consistent with previous isolated and unresolved estimations. Our results demonstrate a new, powerful technique that harnesses the omnipresent nature of sunspot oscillations to provide magnetic field mapping capabilities close to a magnetic source in the solar corona.**

Solar active regions are one of the most magnetically dynamic locations in the Sun's tenuous atmosphere. The intrinsically embedded magnetic fields support a wide variety of coronal compressive MHD wave phenomena, with EUV observations dating back to the late 1990s revealing a wealth of magneto-acoustic oscillations travelling upwards from the solar surface along magnetic fields to distances far exceeding many tens of thousands of km[8-11]. The weak dissipation mechanisms experienced in the solar corona, in the forms of thermal conduction, compressive viscosity and optically thin radiation[12], allow features to be probed through seismology techniques in an attempt to uncover atmospheric changes and sub-resolution structuring that would otherwise be difficult, if not impossible to determine[13]. MHD seismology of sunspots is a powerful tool that has risen in popularity in recent years, with insights into atmospheric structuring, expansion and emission gained through the comparison of wave periods, velocities and dissipation rates[14,15]. Importantly, evidence continues to be brought forward that suggests compressive MHD wave phenomena originating in sunspots is not simply a rare occurrence, but instead an omnipresent feature closely tied to all atmospheric layers[16].

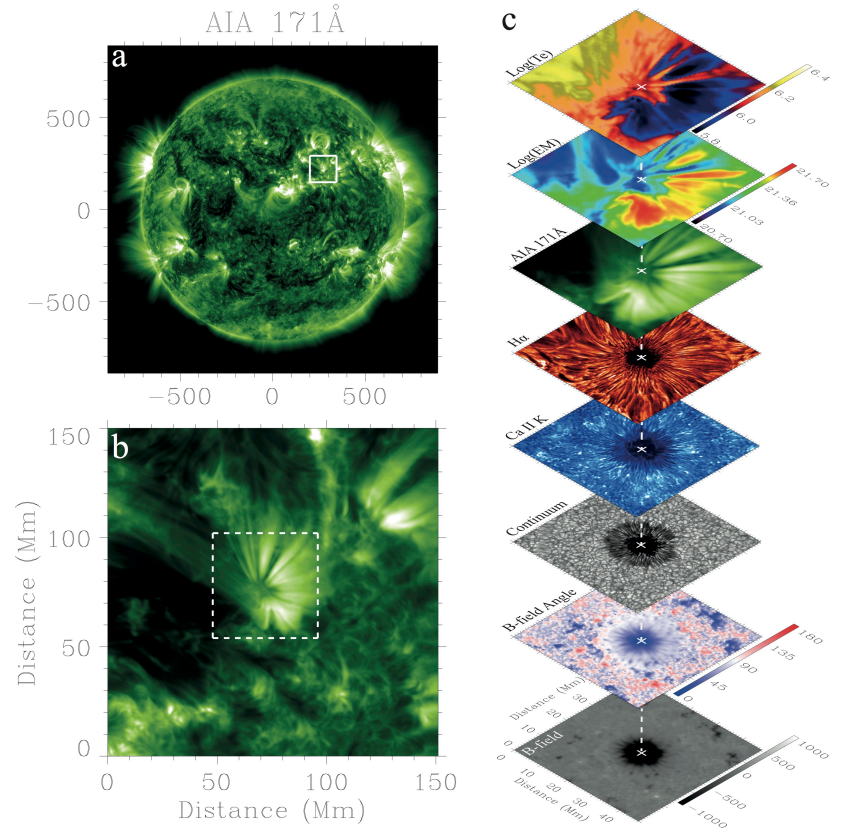

*Figure 1: **The building blocks of the magnetized solar atmosphere observed on 2011 December 10.** (a) A contextual EUV 171Å image of the entire Sun at 16:10 UT, where a solid white box highlights a 150 x 150 Mm² subfield surrounding NOAA 11366. (b) A zoom-in to the subfield outlined in (a), with a further 50 x 50 Mm² subfield highlighted using dashed white lines to show the pointing and size of the ground-based field-of-view. (c) Co-spatial images, stacked from the solar surface (bottom) through to the outer corona (top). From bottom-to-top the images consist of the photospheric magnetic field strength normal to the solar surface ($B_z$; artificially saturated at ±1000 G to aid clarity), the inclination angles of the photospheric magnetic field with respect to the normal to the solar surface (0º and 180º represent magnetic field vectors aligned downwards and upwards, respectively, to the solar surface), ROSA 4170Å continuum intensity, ROSA Ca II K lower chromospheric intensity, HARDcam Hα upper chromospheric intensity, coronal AIA 171Å intensity, emission measure (on a log-scale in units of $cm^{-5}K^{-1}$) and kinetic temperature (displayed on a log-scale artificially saturated between $log(T_e)=5.8$ and $log(T_e)=6.4$ to assist visualisation) snapshots. White crosses mark the location of the umbral barycenter and are interconnected as a function of atmospheric height using a dashed white line.*

Using previously deployed MHD seismology techniques to infer the strength of the coronal magnetic field is a difficult task, often limited by substantial errors (up to 80% in some cases[4]). Critical problems are its inability to deduce the magnetic field strength as a function of spatial position, and by the relative rareness of oscillatory phenomena in the solar corona demonstrating measurable harmonic overtones[7]. Other methods, including deep exposures of coronal magnetic Stokes profiles[17] and the examination of brightness–temperature spectra from gyro-resonance emission[18], also bring about their own distinct impediments, including crosstalk between Stokes signals, poor temporal resolutions often exceeding 1 hour and ambiguous cyclotron frequency reference points[19]. Therefore, while diverse, yet encouraging attempts have been made to constrain the magnitude of the coronal magnetic field in previous years, each process is hindered by its inability to accurately *and* simultaneously deduce spatially resolved, unambiguous and high-cadence field strengths. In this Letter we document and apply a novel MHD-based technique that combines high-resolution observations from modern space-based observatories, vector magnetic field extrapolations and differential emission measure techniques to accurately measure the spatial dependence of the coronal magnetic field strength with potential temporal resolutions orders-of-magnitude better than current imaging approaches.

High spatial (435 km per pixel) and temporal resolution (12 s) observations with the Atmospheric Imaging Assembly[20] (AIA) onboard the Solar Dynamics Observatory (SDO) spacecraft have allowed us to resolve fine-scale and rapidly dynamic coronal features associated with a powerful magnetic sunspot. The data are obtained through six independent EUV filters on 2011 December 10, with the resulting fields-of-view cropped to 50 x 50 Mm$^2$ to closely match simultaneous ground-based observations obtained with the Rapid Oscillations in the Solar Atmosphere[21] (ROSA) and Hydrogen-Alpha Rapid Dynamics camera[14] (HARDcam) instruments on the Dunn Solar Telescope (see supplementary material for further information on the processing of the observational data). The images (Figure 1) show a number of elongated structures extending away from the underlying sunspot, which are particularly apparent in AIA 171Å data that is known to correspond to lower coronal plasma with temperatures ~700,000 K. Differential emission measure (DEM) techniques[22] are employed on the six EUV AIA channels, producing emission measure ($EM$) and temperature ($T_e$) estimates for the sunspot and its locality. The resulting $T_e$ map reveals that the elongated structures, seen to extend away from the underlying sunspot, are comprised of relatively cool coronal material (<1,000,000 K), hence explaining why they demonstrate such visibility in the AIA 171Å channel. Examination of time-lapse movies (Movie S1) reveals a multitude of wave fronts seen to propagate radially outwards from the sunspot, particularly along the relatively cool, elongated coronal fan structures.

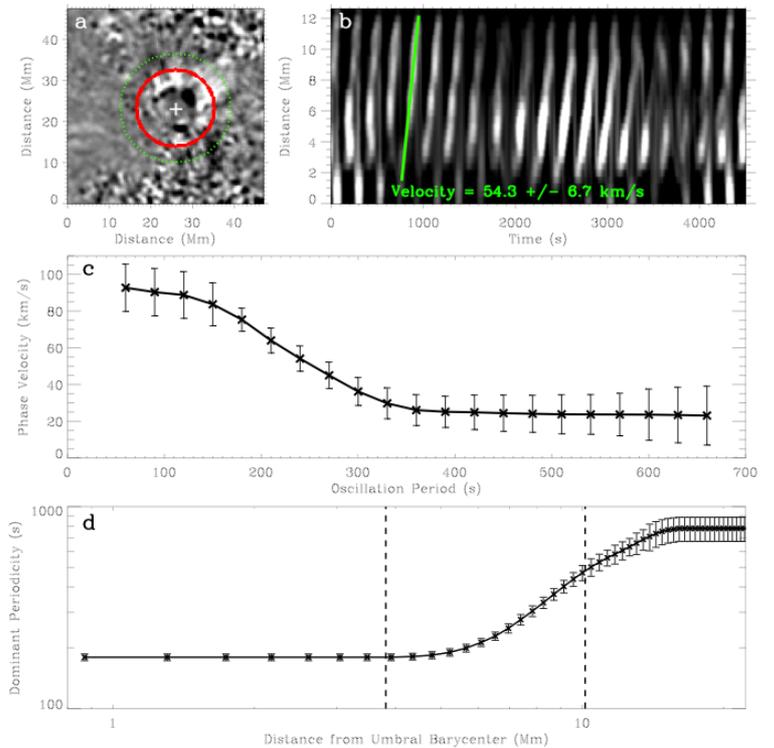

*Figure 2: **Measured velocities and periodicities in the vicinity of a sunspot.** (a) A snapshot of a 171Å intensity image having first been passed through a 4-minute temporal filter. The umbral barycenter is marked with a white cross, while the instantaneous peaks and troughs of propagating waves are revealed as white and black intensities, respectively, immediately surrounding the sunspot. The solid red line highlights an expanding annulus, while the dotted green line outlines the entire spatial extent used in the creation of the time–distance diagram (b), whereby the area delimited by the dotted green curve is azimuthally averaged into a one-dimensional line, directed radially away from the umbral barycenter. (b) The resulting time–distance diagram, consisting of 30 spatial (approximately 13 Mm) by 375 temporal (75 minutes) pixels$^2$. A solid green line highlights a line-of-best-fit used to calculate the period-dependent phase speeds. (c) The period-dependent phase speeds (in km/s), following the Fourier filtering of the time series, plotted as a function of oscillatory period. The error bars show the standard deviations associated with the minimization of the sum of the squares of the residuals in (b). (d) The dominant propagating wave periodicity, measured from normalized and azimuthally-averaged Fourier power maps, as a function of radial distance from the underlying umbral barycenter. The error bars relate to the Fourier frequency resolution of the SDO/AIA instrument operating with a cadence of 12s. The vertical dashed lines represent the umbral and penumbral boundaries established from photospheric SDO/HMI continuum images at approximately 3.8 Mm and 10.1 Mm, respectively, from the center of the sunspot umbra.*

Fourier-filtered images (Figure 2, Movie S2) reveal a wealth of oscillatory signatures covering a wide range of frequencies, with longer periodicities preferentially occurring at increasing distances from the sunspot umbra, and propagating with reduced phase speeds. Importantly, the waves appear to be continuous in time, suggesting that such oscillatory phenomena are driven from a regular, powerful and coherent source. Through center-of-gravity techniques the center of the sunspot (or umbral barycenter) is defined, as indicated by the white crosses in Figure 1. A series of expanding annuli, centered on the umbral barycenter, are used to map the azimuthally-averaged phase velocities as a function of period through the creation of time–distance diagrams (Figure 2). Phase speeds are calculated from the gradients of the diagonal bands in the time–distance diagrams and found to span 92 km/s (60 s periodicity) through to 23 km/s (660 s periodicity), which are characteristic of coronal slow magneto-acoustic tube speeds[9-12], $c_t$. Due to the continuous nature of the propagating waves, Fourier methods are applied to the unfiltered data to create power maps allowing the signatures of the wave phenomena to be easily extracted as a function of period. Employing the same expanding annuli technique, the period-dependent power spectra are studied as a function of the radial distance from the umbral barycenter (Figure S1). A gradual change in the dominant period as a function of distance is found, with periodicities of approximately 3 minutes within the underlying sunspot umbra, increasing rapidly within the penumbral locations, and eventually plateauing at periodicities of approximately 13 minutes further along the coronal fan structures (Figure 2). This suggests that the waves detected within the immediate periphery of the underlying sunspot are the lower coronal counterpart of propagating compressive magneto-acoustic oscillations driven upwards by the underlying sunspot.

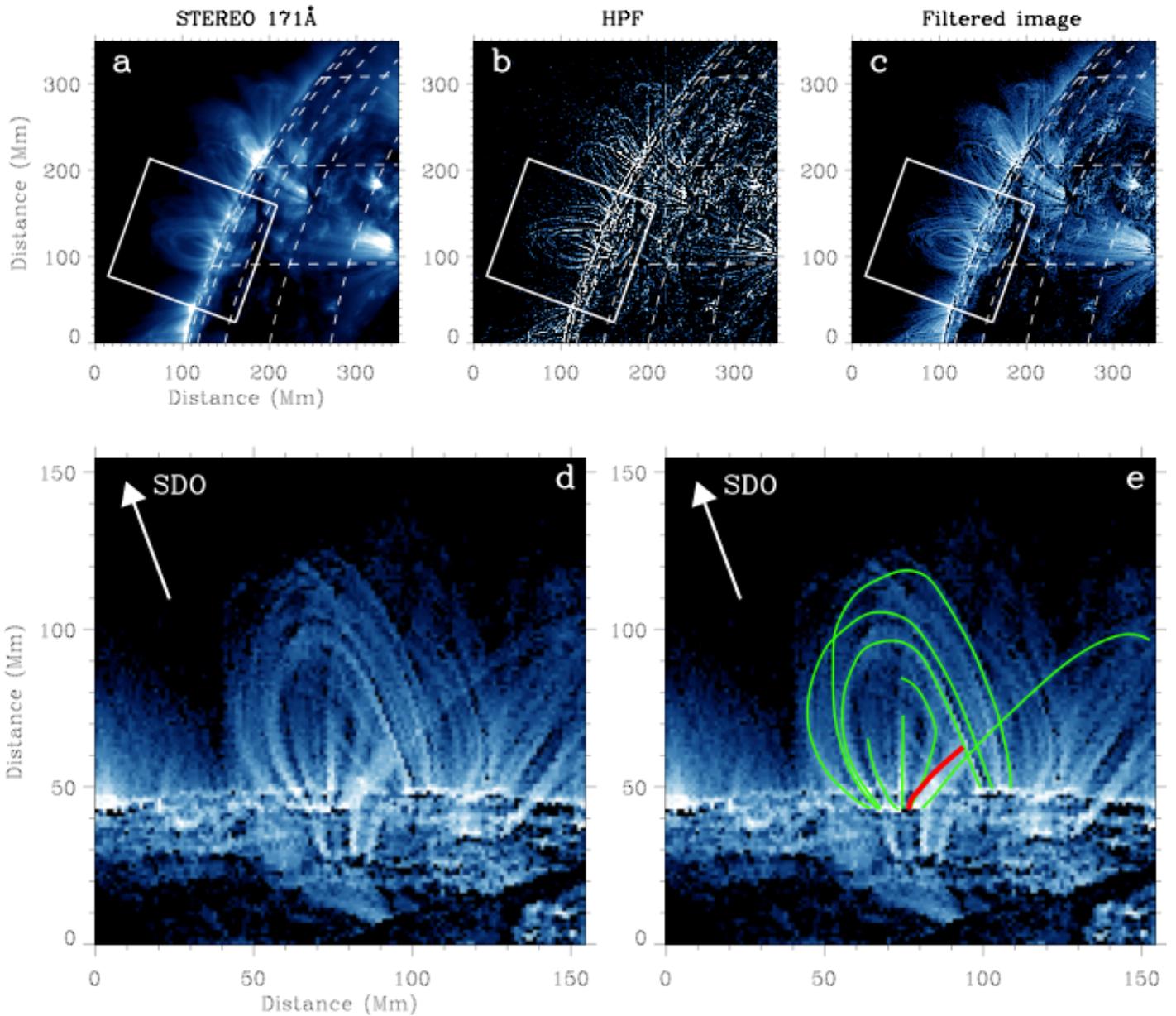

*Figure 3: **A side view of the coronal fan from filtered STEREO–A images.** (a) A 171Å image acquired by the STEREO–A spacecraft at 16:14 UT on 2011 December 10, where white dashed lines display solar heliographic co-ordinates with each vertical and horizontal line separated by 10º. The solid white box highlights the region extracted for measurements in panels **(d)** and **(e)**. **(b)** The STEREO–A image having been ran through a high-pass filter to reveal small-scale coronal structuring. **(c)** The resulting filtered image (i.e. the addition of panels **(a)** and **(b)**) detailing the improved contrast associated with fine-scale coronal loops and fans. **(d)** A zoom-in to the region highlighted by the white boxes in panels **(a–c)** and rotated so the solar limb is horizontal. The white arrow indicates the direction towards the geosynchronous orbit of the SDO spacecraft. **(e)** The same as panel **(d)** only with the addition of traces corresponding to the spatial position of coronal loops in the immediate vicinity of NOAA 11366 (green lines) and the coronal fan currently under investigation (red line).*

Complementary 171Å data from the EUVI instrument onboard the leading Solar TErrestrial RElations Observatory[23] (STEREO–A) provides a side-on view of the coronal fan, allowing the spatial variation of the structure with atmospheric height to be measured directly (Figure 3). Magnetic field extrapolations[24] of vector magnetograms from SDO's Helioseismic and Magnetic Imager[25] (HMI) are also used to calculate magnetic field vectors over a wide range of atmospheric heights, allowing field inclination angles to be compared with direct STEREO–A imaging as a function of distance from the umbral barycenter (Figure 4). We find that increasing the distance from the center of the sunspot results in more inclined magnetic fields. Importantly, this means that the measured phase speeds of the propagating waves will have been underestimated to a certain degree, particularly for locations close to the umbral barycenter where the magnetic field inclination is minimal. To compensate for this, we first use the velocity–period and period–distance relationships to convert the measured tube speeds into a function of distance from the center of the underlying sunspot (Figure 4). Then, the inclination angles at this spatial position are used to calculate the true wave velocity irrespective of inclination-angle effects (Figure 4). Tube speeds of approximately 185 km/s are found towards the center of the umbra, decreasing radially within several Mm to values of approximately 25 km/s.

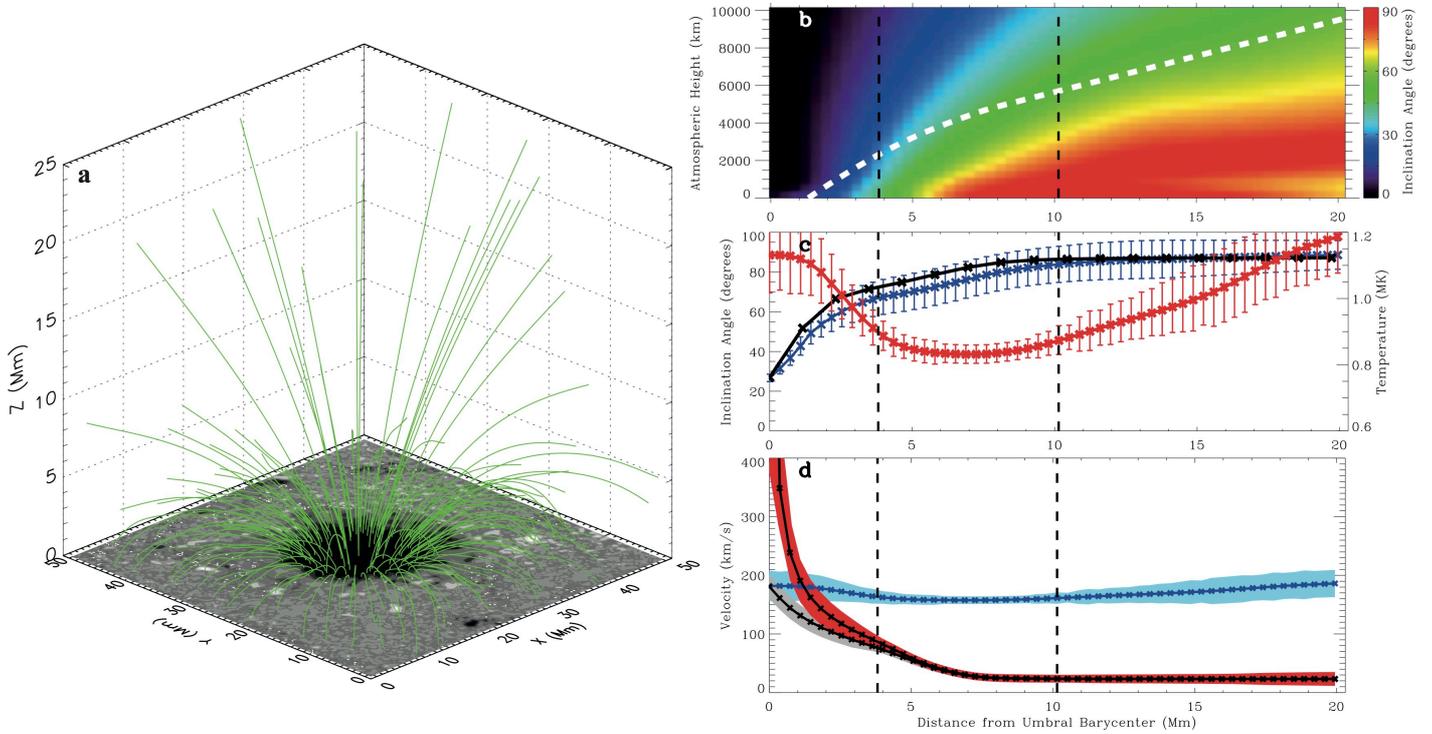

*Figure 4: **Magnetic, temperature and velocity structuring of the sunspot plasma. (a)** The extrapolated magnetic fields (green lines) overlaid on the photospheric $B_z$ map, and viewed from an angle of 45º to the solar surface. **(b)** The azimuthally-averaged inclination angles of the extrapolated magnetic field vectors with respect to the normal to the solar surface, displayed as a function of both the distance from the umbral barycenter and the atmospheric height. Here, 0° and 90° are angles parallel and perpendicular, respectively, to the normal to the solar surface. The dashed white line overlays the coronal fan position as observed by STEREO–A 171Å images. **(c)** The inclination angles of the coronal fan measured from STEREO–A images (black line) alongside the extrapolated magnetic field inclination angles (blue line) corresponding to the dashed white line in **(b)**. Both sets of inclination angles have been transposed into the viewing perspective of the SDO spacecraft by taking into account the location of SDO with respect to the normal to the solar surface. Error bars display the standard deviations associated with the azimuthal averaging of the inclination angles contained within each corresponding annuli. A considerable agreement between the measured and extrapolated field inclination angles demonstrates the accuracy of potential force-free magnetic extrapolations in this instance. The azimuthally-averaged plasma temperature (in MK) derived from DEM techniques is shown using a red line, where the error bars represent the temperature widths associated with DEM fitting techniques. **(d)** The measured tube speed ($c_t$; black line with grey error contours) following compensation from inclination angle effects to provide the true tube speed irrespective of angle ambiguities, the temperature-dependent sound speed ($c_s$; blue line with light blue error contours) calculated from the local plasma temperature, and the subsequently derived Alfvén speed ($v_A$; black line with red error contours) displayed as a function of distance from the underlying umbral barycenter. Errors in the sound speeds and tube speeds are propagated from the temperature width and a combination of inclination angle **(c)** plus Fourier frequency resolution (Figure 2**(d)**) plus residual phase velocity measurement (Figure 2**(c)**) errors, respectively. The errors associated with the Alfvén speeds result from the combination of sound speed and tube speed error estimates. The vertical dashed lines represent the umbral and penumbral boundaries established from photospheric SDO/HMI continuum images at approximately 3.8 Mm and 10.1 Mm, respectively, from the center of the sunspot umbra.*

Since the sound speed, $c_s$, is dependent upon the local plasma temperature, we use the $T_e$ maps to derive the spatial distribution of the sound speed through the relation,

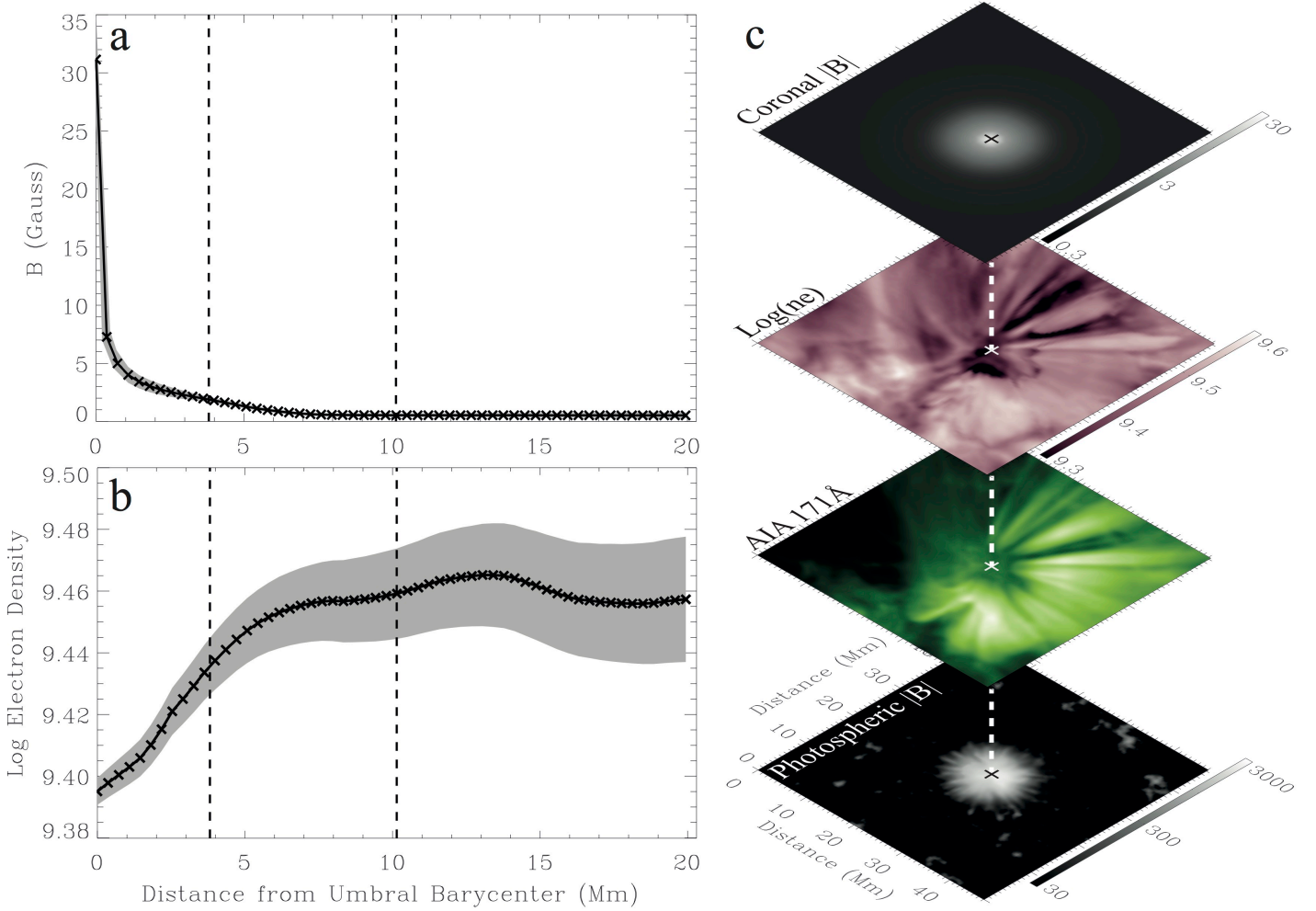

*Figure 5: **The magnitude of the magnetic field in the solar corona.** (a) The magnitude of the magnetic field strength, |B|, plotted as a function of radial distance from the umbral barycenter, where the grey error shading represents the propagation of errors associated with the Alfvén speeds (Figure 4(d)) and the plasma densities (b). (b) The azimuthally-averaged electron densities, $n_e$, displayed on a log-scale (units of $cm^{-3}$) using the same distance scale as (a), where the vertical dashed lines represent the umbral and penumbral boundaries established from photospheric SDO/HMI continuum images at approximately 3.8 Mm and 10.1 Mm, respectively, from the center of the sunspot umbra. Greyscale error shading highlights the goodness of fit that is defined by a least-squares optimization of the plasma emission measures using DEM approaches. (c) Co-spatial images providing a two-dimensional representation of the magnitude of the photospheric magnetic field strength (lower; displayed on a log-scale and saturated between 30 and 3000 G to aid clarity), the EUV AIA 171Å intensity (middle lower), the electron densities (middle upper; saturated between $log(n_e)$=9.3 and $log(n_e)$=9.6 to assist visualization) and the reconstructed magnitude of the coronal magnetic field strength (upper; displayed circularly symmetric on a log-scale and saturated between 0.3 and 30 G to aid clarity). White and black crosses indicate the position of the umbral barycenter, which are connected between atmospheric heights using a dashed white line.*

$$c_s = \sqrt{\gamma R T_e / \mu}$$

where γ=5/3 is the adiabatic index, µ=1.3 g/mol is the mean molecular weight assuming H:He=10:1 for the solar corona[26], and R is the gas constant. This provides sound speeds larger than the compensated tube speeds, and of the order of 160–210 km/s (Figure 4). With the distributions of $c_s$ and $c_t$ known, we estimate the spatial distribution of the Alfvén speed, $v_A$, through the relationship[27],

$$v_A = \frac{c_s c_t}{\sqrt{c_s^2 - c_t^2}}$$

The derived Alfvén speeds, displayed in Figure 4, are highest (1500 km/s) towards the center of the sunspot umbra, dropping to their lowest values (25 km/s) at the outermost extremity of the active region. The large values found towards the sunspot core are consistent with previous coronal fast-mode wave observations related to magnetically confined structures[28].

The Alfvén speed depends on two key parameters: the magnitude of the magnetic field strength, $|B|$, and the local plasma density, $\rho$. We calculate[22] spatially-resolved plasma densities from the $EM$ and $T_e$ maps, resulting in a range of densities on the order of $(2.1–5.6) \times 10^{-12}$ kg/m$^3$. With the Alfvén speeds and local plasma densities known, we derive the magnitude of the azimuthally averaged magnetic field at each radial location through the relationship[27],

$$|B| = v_A \sqrt{\mu_0 \rho}$$

where $\mu_0$ is the magnetic permeability. The resulting values for $|B|$ range from 32 ± 5 G at the center of the underlying sunspot, rapidly decreasing to 1 G over a lateral distance of approximately 7000 km (Figure 5). Both our maximal and minimal values are consistent with previous, yet isolated and/or error-prone independent estimations of magnetically dominated coronal loops[4-7] and quiet Sun locations[29]. However, importantly, we show for the first time a novel way of harnessing the omnipresent nature of propagating slow-mode waves in the corona to more-accurately constrain the magnetic field topology as a function of spatial location, with the potential to provide coronal $B$-field maps with cadences as high as 1 minute. This would provide more than an order-of-magnitude improvement in temporal resolution compared with deep exposures of coronal Stokes profiles[17]. A key science goal of the upcoming 4m Daniel K. Inouye Solar Telescope (DKIST) is to directly image the full coronal Stokes profiles using near-infrared spectral lines. The novel techniques presented here will therefore serve as an important long-term tool that can be used to compare directly with the DKIST observations following first light in 2019. Furthermore, since magnetic reconnection phenomena are often observed in the vicinity of active regions, a high-cadence approach to monitoring the spatial variations of the coronal magnetic field would be critical when attempting to understand the pre-cursor events leading to solar flares.

**Acknowledgements:** D.B.J. wishes to thank the UK Science and Technology Facilities Council (STFC) for the award of an Ernest Rutherford Fellowship in addition to a dedicated Research Grant. D.B.J. also wishes to thank Invest NI for their Research & Development Grants. This work was supported by a UKIERI trilateral research grant of The British Council.

**Author Contributions:** D.B.J., V.E.R., S.K.P., S.Y. and C.D. performed analysis of observations. D.B.J., M.M., S.K.P. and D.B. interpreted the observations. D.B.J., P.H.K., S.Y., R.S.I.R., D.J.C., S.D.T.G. and C.D. performed all image processing. D.B.J., D.H.M., D.J.C. and M.M. designed the observing run. All authors discussed the results and commented on the manuscript.

## Supplementary Information:

**Data overview and requirements:**
The current Letter employs a vast array of ground- and space-based observations, in addition to the application of modern magnetic field extrapolation algorithms, to detail the study of magnetic fields in the Sun's corona. However, it must be stressed that not all simultaneous and complementary datasets are required to fully exploit the techniques presented here. For example, Figure 1 displays cotemporaneous images from both ground-based optical instrumentation and space-borne EUV satellites in order to allow the reader to fully visualize the connectivity of the sunspot from the surface of the Sun (i.e., the photosphere) through to the outermost extremities of the tenuous corona. Thus, while the ground-based optical images help to represent the atmospheric structuring of the sunspot for contextual purposes, the techniques and results documented here originate solely from data obtained by user-friendly space-based facilities. Furthermore, in order to examine the inclination angles associated with the magnetic structures in the Sun's corona we employ both a non-linear force-free field extrapolation code[24], in addition to utilizing simultaneous EUV image sequences obtained with the STEREO spacecraft. Both methods are documented to highlight the excellent agreement between modern magnetic field extrapolation codes and observations from stereographic imaging instruments. As a result, the techniques and results presented here only require a single complementary measurement (e.g., from an alternative viewing perspective such as the STEREO spacecraft or by employing widely available magnetic field extrapolation algorithms).

We would therefore like to stress that the widespread use of complementary datasets throughout this Letter is solely to validate the techniques applied, and ultimately verify the results we present. Future applications of the processes documented here can be applied directly to continuous observations obtained by the Solar Dynamics Observatory[3S], in addition to either widespread magnetic field extrapolation codes or near-simultaneous imaging sequences acquired from an alternative viewing perspective (e.g., STEREO or the upcoming Solar Orbiter).

**Observational data processing:**
The Helioseismic and Magnetic Imager[25] (HMI) onboard the Solar Dynamics Observatory[3S] (SDO) is employed to provide Milne-Eddington vector magnetograms of active region NOAA 11366, in addition to contextual photospheric continuum snapshots. The HMI data is processed using the standard 'hmi_prep' SSWIDL routine, with 200 arcsec x 200 arcsec (145 x 145 $Mm^2$) subfields surrounding the sunspot extracted with a cadence of 720 s and a two-pixel spatial resolution of 1.0 arcsec (725 km). To complement the high-resolution optical imaging and vector magnetogram datasets, coronal EUV image sequences are obtained by the Atmospheric Imaging Assembly[20] (AIA) onboard SDO with a two-pixel spatial resolution of 1.2 arcsec (870 km) and a temporal cadence of 12 s. Six independent EUV bandpasses are processed using the standard 'aia_prep' SSWIDL routine, consisting of the 94Å, 131Å, 171Å, 193Å, 211Å and 335Å channels, with the data spanning a full 75 minutes, providing 375 independent snapshots per bandpass of NOAA 11366, with typical effective temperatures of approximately 7.0 MK, 0.4 MK, 0.7 MK, 1.6 MK, 2.0 MK and 2.8 MK, respectively[4S].

The EUVI instrument onboard the Solar TErrestrial RElations Observatory[23] (STEREO) is employed to provide 171Å images of NOAA 11366 from a different vantage point to that provided by SDO. On 2011 December 10 the leading STEREO–A spacecraft was approximately 107º ahead of the Earth in its orbit around the Sun, while the STEREO–B satellite trailed the Earth by approximately 108º. Due to the westerly positioning (as observed by SDO) of NOAA 11366, it was only visible by the leading STEREO–A spacecraft as an active region positioned very close to the solar limb (see Figure 3). The STEREO–A instruments were positioned 0.062º below the Sun–Earth ecliptic[5S], and therefore present an ideal way of comparing well-aligned (in the N–S plane)

171Å images with those obtained from the geosynchronous orbit of SDO. During the observing sequence presented here, the EUVI instrument only obtained a single 171Å image at 16:14 UT with a two-pixel spatial resolution of 3.2 arcsec (2320 km). This image is prepared using the standard 'secchi_prep' SSWIDL routine, before being cropped to a 300 x 300 Mm$^2$ subfield surrounding active region NOAA 11366 for subsequent study.

The ground-based observational data presented here was obtained during 16:10 – 17:25 UT on 2011 December 10, with the Dunn Solar Telescope (DST) at Sacramento Peak, New Mexico. While the preparation of ground-based image sequences are described here, with the resulting snapshots used for contextual purposes in Figure 1, it is the images and vector magnetograms from space-based observatories that form the main focus of the Letter. The Rapid Oscillations in the Solar Atmosphere[21] (ROSA) and Hydrogen-Alpha Rapid Dynamics camera[14] (HARDcam) ground-based, multi-wavelength imaging systems are employed to image a nearly circularly symmetric sunspot within active region NOAA 11366, close to disk center with heliocentric co-ordinates (356 arcsec, 305 arcsec), or N17.9W22.5 in the conventional heliographic co-ordinate system. ROSA observations are acquired through a 52Å bandpass continuum filter centered at 4170Å, in addition to a 1Å bandpass Ca II K filter centered at 3933Å. Each filtergram employs a common plate scale of 0.069 arcsec per pixel (50 km per pixel), providing a diffraction-limited field-of-view size of approximately 69 arcsec x 69 arcsec (50 x 50 Mm$^2$). HARDcam observations employ a narrowband filter centered on the H$\alpha$ line core (6562.8Å), and utilise a spatial sampling of 0.138 arcsec per pixel (100 km per pixel) to provide a field-of-view size (71 arcsec x 71 arcsec; 51 x 51 Mm$^2$) comparable to the ROSA continuum image sequence. During the observations, high-order adaptive optics[1S] are used to correct wavefront deformations in real-time with the images further improved through speckle reconstruction algorithms[2S]. The resulting post-reconstruction cadences for the continuum, Ca II K and H$\alpha$ images are 2.11 s, 4.22 s and 1.78 s, respectively. Atmospheric seeing conditions remained excellent throughout the time series. However, to ensure accurate co-alignment between the bandpasses, broadband time series are Fourier co-registered and de-stretched using a 40x40 grid, equating to a 1.7 arcsec separation between spatial samples. Contextual HMI (6173Å) and AIA (4500Å) continuum images, acquired at 16:10 UT, are obtained for the purposes of co-aligning the SDO data with images of the lower solar atmosphere. Using the AIA continuum context image to define absolute solar co-ordinates, the HMI and ground-based observations are subjected to cross-correlation techniques to provide sub-pixel co-alignment accuracy between the imaging sequences. To do this, the plate scales of both HMI and ground-based observations are first degraded to match that of the AIA continuum image (note that subsequent data analysis is performed on full-resolution (i.e., non-degraded) image sequences). Next, squared mean absolute deviations are calculated between the datasets, with the HMI and ground-based images subsequently shifted to best align with the AIA reference image. Uncertainties are estimated as less than one tenth of a pixel of the coarsest resolution instrument (i.e., 0.06 arcsec or 43.5 km based on the spatial sampling of AIA).

**Computing the $T_e$ and *EM* maps:**
Differential emission measure (DEM) techniques[22] are employed in such a way to ensure accurate co-alignment between different AIA wavelength images, thus providing reliable and accurate $T_e$ and *EM* mapping capabilities. Measurements of the altitude of the EUV-absorbing chromospheres are computed to ensure pointing offsets between the EUV bandpasses are minimal (errors on the order of one tenth of an AIA pixel, or 45 km). In addition, each *EM* and $T_e$ map is self-calibrated through the use of empirically corrected instrumental response functions of the AIA imagers, thus providing accurate measurements of the optically thin solar corona. Since the sunspot and surrounding active region remained inactive (i.e., no large scale eruptive phenomena detected) during the 75 minute observing sequence, only single point DEM methods are required (i.e., no double fitting of non-isothermal plasma). The resulting output maps are complemented by estimates of the temperature width, in addition to the goodness of fit that is defined by a least-squares optimization, providing error estimates for the derived temperatures and emission measures[22]. Furthermore, due to the continuous presence of the coronal fan and loop structures in the 75-minute duration of the dataset, it is beneficial to temporally average the AIA image sequences prior to implementing the DEM algorithms in order to improve the resulting signal-to-noise. Thus the time-averaged $T_e$ and $EM$ maps provide a more general overview, devoid of small-scale and/or rapidly moving/evolving transients, and thus best represent the basal characteristics of the sunspot, including its immediate vicinity.

**Fourier regeneration of time series:**
To verify the presence of wave phenomena in the AIA 171Å field-of-view, we employ a Fourier-based filtering algorithm[16] to isolate and re-generate new time series which had been decomposed into frequency bins, each separated by 30 s, and corresponding to periodicities of 1–11 minutes. This allows propagating periodic phenomena to be more easily identified and quantified since it removes the superposition of numerous frequencies and phase speeds within localised regions. Each Fourier filter bin is peaked at 60, 90, 120, …, 660 s, and incorporates a Gaussian-profiled window spanning a half-width half-maximum of ±15 s. The Fourier filtering algorithm acts only as a frequency filter and does not make any assumptions regarding the spatial scales associated with a particular frequency, and as a result performs no filtering on the spatial wavenumber (i.e., $k$).

**Defining the umbral barycenter:**
The location of the center of the sunspot is found using a 'center-of-gravity' technique. All of the contextual SDO/HMI continuum images (100 in total) are time-averaged to produce an intensity map devoid of small-scale and short-lived features. Next, the umbral perimeter is defined as the boundary where pixel intensities decreased below 45% of the median granulation intensity. Features brighter than this are discarded, producing an accurately defined outline of the sunspot umbra, containing approximately 250 pixels and encompassing an area of $3.3 \times 10^7$ km$^2$. From this point, the umbral center-of-gravity, or intensity 'barycenter', is established, which forms the central co-ordinates of the expanding annuli used in subsequent analysis.

In order to validate the accuracy of calculating the umbral barycenter from SDO/HMI continuum images, we also compute the center of the sunspot using the higher spatial resolution 4170Å continuum observations obtained by ROSA. As with the SDO/HMI observations, all consecutive 4170Å continuum images (2130 in total) are time-averaged to produce an intensity map devoid of small-scale and short-lived features that are particularly apparent in high-resolution datasets (e.g., umbral dots, magnetic bright points, and to a certain extent, granulation). The umbral perimeter is then defined as the boundary where pixel intensities decreased below 45% of the median granulation intensity, producing an accurately contoured umbral perimeter containing approximately 15,000 pixels (nearly two orders-of-magnitude more than the simultaneous SDO/HMI images). Applying the center-of-gravity technique once again defined the umbral barycenter, which was found to lie within the same resolution element as the co-ordinates calculated directly from SDO/HMI continuum images. Thus, the umbral barycenter derived from SDO/HMI images can be considered accurate, especially since cotemporaneous SDO/AIA and STEREO image sequences, to which the umbral barycenter will be used in conjunction with, have more coarse spatial resolution than SDO/HMI.

**Creating the expanding-annuli series:**
Previously[16], an expanding series of circular annuli were employed to measure the azimuthally-averaged radial variations of running penumbral wave phenomena in the solar chromosphere as a function of distance from the umbral barycenter. This was deemed suitable in the lower solar atmosphere where the underlying sunspot displayed symmetric structuring in most radial directions. In order to make the techniques documented here as widely applicable as possible, we also implement an expanding series of circular annuli to analyze the prominent propagating wave phenomena detected in SDO/AIA image sequences. Examination of the $T_e$, *EM* and AIA 171Å intensity maps clearly reveals that the coronal plasma in the immediate viscinity of the sunspot is relatively cool, with basal temperatures on the order of 600,000 – 1,200,000 K (Figure 1). Due to these temperatures being optimally placed within the response curve of the 171Å channel[4S], and coupled with its optically thin characteristics, high signal-to-noise values, and due to the resulting images residing within the lower, cooler corona, the 171Å bandpass was deemed ideal for subsequent study. We therefore utilize a series of expanding annuli to more readily measure coronal variations as a function of radial distance from the umbral barycenter. Each annulus is 2 pixels wide (870 km), with subsequent annuli spaced further away by 1 pixel from the preceding measurement location. Some overlap between adjacent annuli is chosen to provide continuous radial coverage of the measured parameters, while still maintaining high pixel numbers to improve statistics. A sample annulus, which is 20 AIA pixels (9 Mm) from the umbral barycenter, is displayed in Figure 2.

For completeness, it must be reiterated that the 171Å channel was selected for the present study due to its response function spanning all derived temperatures within the sunspot locality and its immediate periphery (see, e.g., Figure 1). However, it must be noted that not all coronal fans and loops display similarly distributed temperatures, with some displaying cooler, hotter or both temperatures within the sunspot vicinity. As it is imperative to promote high signal-to-noise values when extracting (often faint) wave signals, future users of these techniques must identify which AIA channels best sample the thermal properties of the loop/fan system under investigation. The associated sound speeds will be pre-determined by the best fitting isothermal approximation (see **Computing the $T_e$ and EM maps** above), thus in order to ensure that the most reliable comparison is made to the local sound speed, the measured wave tube speeds must be extracted from data obtained by an AIA channel that closely matches the local coronal temperature. With typical AIA response functions spanning 0.4 – 7.0 MK[4S], including those that can be sub-divided yet further into multiple temperature peaks[6S], almost all perceivable solar active regions and loop configurations can be investigated using the techniques presented in this Letter when accompanied by an AIA imaging channel that best represents the local coronal temperature profile.

**Measuring the tube speeds from time–distance diagrams:**
To quantify the propagating phase speed characteristics visible within the Fourier-filtered time series, time–distance analysis is employed with the starting point positioned at the center of the underlying sunspot (white cross in Figure 2). The expanding series of annuli (see **Creating the expanding-annuli series** above) are used to isolate intensities as a function of time and distance from the umbral barycenter. The intensities falling within a single annulus are azimuthally averaged to generate a time–distance diagram that only depends on the radial distance from the center of the sunspot and the time when the measurement is acquired, hence creating a time–distance diagram similar to that displayed in Figure 2. A sample annulus is displayed as a solid red line surrounding the sunspot depicted in Figure 2. The propagating oscillatory signatures, visible as recurring white and black diagonal bands, importantly show that the waves appear to be continuous in time, suggesting that such oscillatory phenomena are driven from a regular, powerful and coherent source. Furthermore, the gradient of the diagonal bands provides an indication of the tube speed, $c_t$, normal to the observer's line-of-sight. These are measured by first fitting a Gaussian profile across the widths of the diagonal bands[7S], before fitting a line-of-best-fit to the resulting Gaussian peaks and minimizing the sum of the squares of the residuals (i.e., least squares fitting). The errors associated with these measurements are larger at very low and very high periods as a result of short propagation distances and reduced signals, respectively, found in the immediate vicinity of the sunspot. Nevertheless, a clear trend is displayed in Figure 2, with tube speeds of approximately 92.3 km/s found for 60 s periodicities, decreasing to 23.2 km/s for 660 s periodicities.

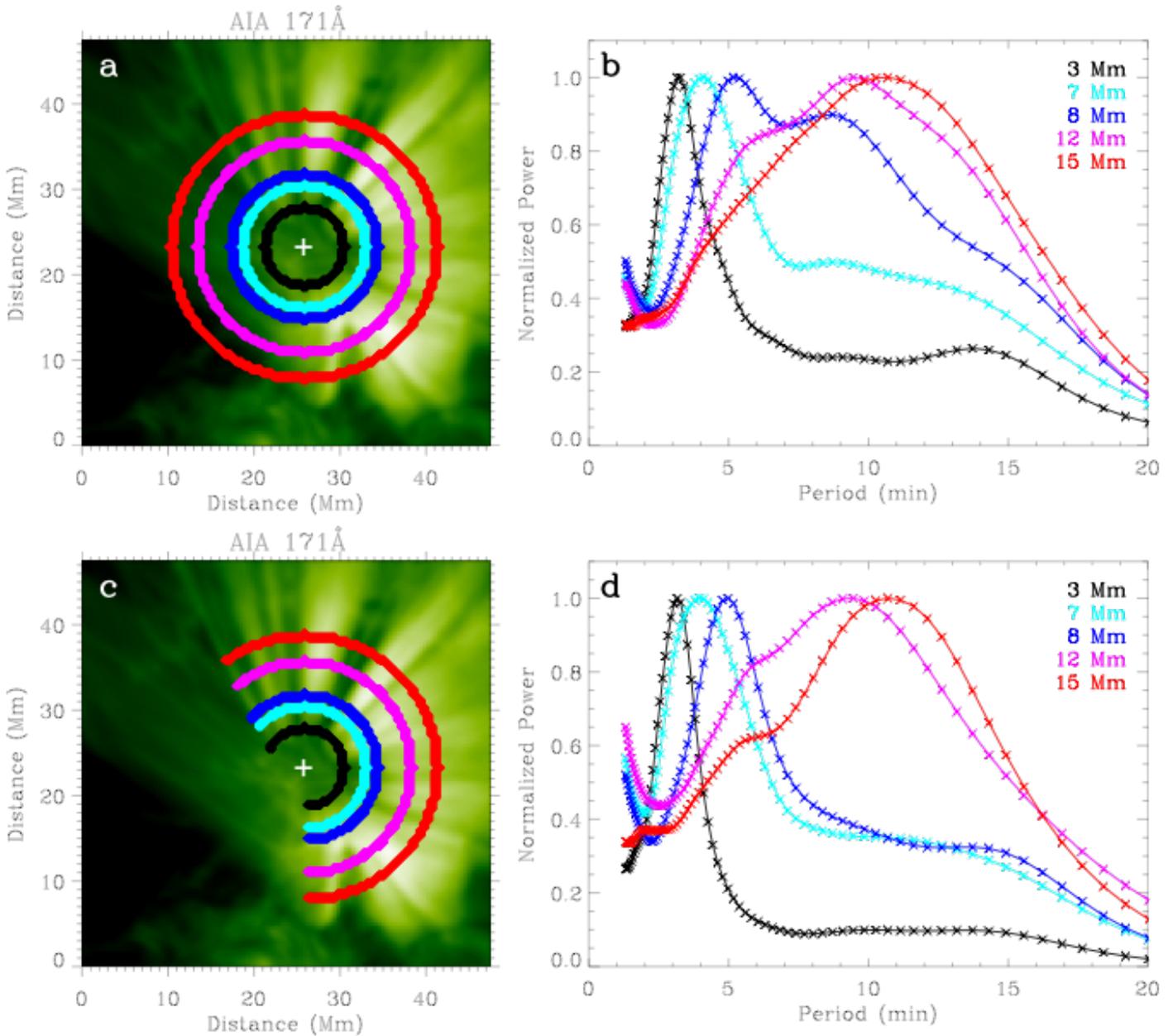

*Figure S1: **The evolution of wave power spectra with increasing distance from the sunspot umbra.** (a&c) An AIA 171Å intensity image overplotted with black, light blue, dark blue, pink and red contours depicting circular annuli (a) and arc segments (b) corresponding to lateral distances of 3 Mm, 7 Mm, 8 Mm, 12 Mm and 15 Mm, respectively, from the umbral barycenter. Fourier spectra obtained within each circular/arc segment are spatially averaged and displayed in **(b&d)**. Using the same color scale as **(a&c)**, **(b&d)** display the resulting spatially averaged power spectra for the five individually chosen circular/arc segments, each normalized to their own respective maximum, as a function of oscillation period. It is clear to see that the dominant period becomes larger (in both cases) as the distance from the sunspot umbra increases.*

**Interpreting the waves:**
Figures 2 and S1 reveal how dominant periodicities of approximately 3 minutes are found within the sunspot umbra, increasing rapidly within the penumbral locations, eventually plateauing at periodicities equal to approximately 13 minutes further along the coronal fan structures. Running penumbral waves[8S,9S] (typically referred to as RPWs) are often most visible in the solar chromosphere as wave fronts propagating outwards from dark sunspot umbrae along the rapidly inclining magnetic fields associated with penumbral and near-sunspot locations. These compressive magneto-acoustic waves are also found to be continuously generated throughout the lifetime of a sunspot, with the chromospheric frequencies and phase speeds largest (>10 mHz or <100 s) at the inner penumbral boundary, decreasing to their lowest values (<1 mHz or >1000 s) towards the outer penumbral edge[16,10S,11S]. Additionally, evidence[16,12S] is documented that such compressive magneto-acoustic oscillations can propagate to distances far exceeding 15 arcsec

(10 Mm) from the outer edge of the sunspot boundary, suggesting the wave properties have significantly strong signals at radial distances well-beyond the confines of the visible sunspot. Thus, a combination of the measured velocity–period and period–distance dependencies, the azimuthally coherent nature of the propagating wave trains, the locations in which the waves are detectable, the sub-sonic phase speeds and the magnetically-confined wave characteristics seem to suggest that we are observing the lower coronal counterpart of propagating compressive magneto-acoustic oscillations driven upwards by the underlying sunspot. This also substantiates the theory that significantly extended propagating wave trains in coronal fans and plumes are simply the observational signatures of large-amplitude umbral oscillations guided upwards along inclined magnetic field lines[8-13].

**Converting velocity–period to velocity–distance:**
Calculating the gradients present in the Fourier-filtered time–distance diagrams allows the phase speeds to be calculated as a function of oscillatory period (i.e., the velocity–period relationship; middle panel of Figure 2). We find phase speeds that span 92 km/s (60 s periodicity) through to 23 km/s (660 s periodicity), which are characteristic of coronal slow magneto-acoustic tube speeds[9-12], $c_t$. Then, employing Fourier analysis on the unfiltered AIA 171Å image sequence allows power maps to be generated over a wide range of periodicities (60–660 s). Following the normalisation of each power spectra by the average power contained within the entire period-specific field-of-view, the dominant periodicity is then defined as the periodicity which has the most relative azimuthally-averaged power within each circular annulus. As found for dedicated chromospheric wave phenomena[16], the 171Å coronal dataset is also found to display a gradual change in the dominant period as a function of distance from the umbral barycenter (i.e., the period–distance relationship; Figures 2 and S1). We find periodicities of approximately 3 minutes within the underlying sunspot umbra, increasing rapidly within the sunspot penumbra, and eventually plateauing at periodicities of approximately 13 minutes further along the coronal fan structures. Since the dominant periodicity is simply a selection of which period displays the highest power signal in a given annulus, the error associated with this period is related directly to the Fourier frequency resolution. The AIA instrument, with an imaging cadence of 12 s, provides excellent high-frequency resolution, as indicated by the error bars in the lower panel of Figure 2. To convert the measured velocity–period agreement into a velocity–distance relationship we utilise the period–distance trend to relate a particular oscillation period to a specific distance from the underlying sunspot umbra, thus creating a direct relationship between the measured wave velocities and the distance from the umbral barycenter (i.e., the velocity–distance relationship). These tube speeds, as a function of distance from the umbral barycenter, are displayed in the lower panel of Figure 4 following the inclination-angle compensation (see below).

**Compensating for inclination-angle effects in the $c_t$ measurements:**
Magnetic field extrapolations are useful to estimate the magnetic topology of the solar atmosphere at locations away from the photosphere. The non-linear force-free field extrapolation code[24] has proved to be accurate and reliable[16] within the magnetic field configuration of active region NOAA 11366. We employ the same annulus-based analysis on the extrapolated magnetic field vectors parallel ($B_x$, $B_y$) and normal ($B_z$) to the solar surface to build up an azimuthally averaged distribution of the inclination angles as a function of both the distance from the umbral barycenter and the atmospheric height (upper panel of Figure 4). It is clear that increasing distance from the center of the sunspot results in more inclined (with respect to the normal to the solar surface) magnetic fields. While the magnetic field vectors parallel and normal to the solar surface can be estimated for coronal heights based upon the field extrapolations, they unfortunately do not provide critical information related to the orientation of the magnetic fields with respect to the observer's line-of-sight. To address this issue, we inversely process[13S,14S] the extrapolated and disambiguated magnetic field vectors to obtain the line-of-sight, east-west and north-south components of the magnetic field with respect to the geosynchronous line-of-sight of the SDO spacecraft. The resulting inclination angles means that the measured tube speeds of the propagating waves will have been underestimated to a certain degree, particularly for locations close to the umbral barycenter where the magnetic field inclination is minimal.

To compensate for this, we are able to utilize the azimuthally averaged magnetic field inclination angles produced by the field extrapolation code, once they have been processed back into the reference frame of the geosynchronous SDO spacecraft. However, in order to verify that this approach is indeed highly accurate and reliable, we also take advantage of the side-on viewpoint of the STEREO–A observations to build up a picture of how the coronal fan height varies as a function of distance from the underlying sunspot. A number of coronal loop structures are also outlined for contextual purposes (green lines in the lower-right panel of Figure 3). While the observations may have much coarser resolution when compared to the AIA onboard SDO, a progressive inclination of the coronal fan is clear with increasing distance from the sunspot (red line in the lower-right panel of Figure 3). With the relative positioning of the instruments known, the inclination angles are transformed into the true reference frame of the SDO spacecraft (black line in the middle panel of Figure 4). To compare directly with the magnetic field extrapolations, the measured height of the coronal fan from STEREO–A observations is overplotted on the extrapolated and azimuthally averaged angle distribution (dashed white line in the upper panel of Figure 4). It is clear that the observed coronal fan positioning closely follows the extrapolated field lines, hence highlighting the validity of using either magnetic field extrapolations or near-simultaneous observations from an alternative vantage point to quantify the inclination angles of the coronal magnetic fields. Then, the inclination angles corresponding to the dashed white line are extracted from the distribution, transformed into the reference frame of the SDO spacecraft, and compared with those directly measured from the STEREO–A observations (blue line in the middle panel of Figure 4). The close agreement between the numerically extrapolated and observationally measured field lines reiterates the accuracy and reliability of the non-linear force-free field extrapolation code within the magnetic field configuration of active region NOAA 11366. It must be noted that two independent methods are applied here to quantify the coronal magnetic field inclination angles. However, future applications of this technique need only to employ a single approach, thus simplifying the process further.

The specific inclination angles of the magnetic fields with respect to SDO's line-of-sight are approximately 30º near the center of the umbra, decreasing to approximately 90º at the outer periphery of the active region. Due to the smooth variation of the measured inclination angles with distance from the sunspot, they are interpolated from the STEREO–A spatial sampling (1.6 arcsec per pixel) onto an AIA grid (0.6 arcsec per pixel) in order to allow direct comparisons between the imaging data. With the measured tube speeds now available as a function of distance from the umbral barycenter (see the above **Converting velocity–period to velocity–distance** section), each measured phase velocity is divided by the co-spatial cos(θ) term (where θ is the inclination angle with respect to SDO's line-of-sight) to convert each value into the true tube speed irrespective of inclination-angle effects. Figure 4 displays the true wave speeds as a function of distance from the umbral barycenter, with velocities now approaching 185 km/s towards the center of the umbra, and decreasing radially within several Mm to values of approximately 25 km/s.

**Calculating the local plasma densities:**
The Alfvén speed depends on two key parameters: the magnitude of the magnetic field strength, $|B|$, and the local plasma density, $\rho$. We are able to calculate spatially-resolved coronal electron densities, $n_e$, from a combination of the *EM* and $T_e$ maps previously derived through DEM techniques[22]. To do this it is necessary to estimate the depth of the coronal fan structure. Previous two- and three-dimensional coronal reconstructions[15S,16S] have demonstrated how the depth of coronal fan structures are approximately 1300 km (or 3 AIA pixels), with minimal scatter. However, since we have simultaneous STEREO–A imaging we are able to measure the thickness of the fan structures directly. Here, our measurements also corroborate recent statistical findings[15S,16S], with a thickness on the order of 1–2 pixels (1160–2320 km; Figure 3). As a result, we adopt the statistically significant depth of 1300 km, which is also consistent with our direct STEREO–A observations.

Employing the *EM* and $T_e$ maps alongside the 1300 km coronal fan depth, we are able to calculate[22] the spatially resolved $n_e$ distribution (Figure 5). The range of electron densities, on the order of Log($n_e$) = 9.2 – 9.6 (1.7x10$^9$ – 3.9x10$^9$ cm$^{-3}$), is consistent with a wide variety of previous coronal density-sensitive line intensity ratios measured using high-resolution spectroscopic instrumentation[17S,18S]. Then, employing the expanding-annulus method on the derived $n_e$ map allows changes in the azimuthally averaged electron densities to be established as a function of distance from the umbral barycenter, with the resulting values plotted in Figure 5. Associated errors in the electron densities are computed from the goodness of fit of the individual DEMs, each of which is defined by a least-squares optimization[22]. As a result of averaging over extended annuli at larger radial distances from the umbral barycenter, the associated electron density uncertainties are observed to increase in magnitude (Figure 5b). Since the 171Å channel is dominated by Fe IX emission at temperatures slightly below 1,000,000 K, it is most likely fully ionised plasma with a ratio of hydrogen mass density to electron density (i.e., $n_H/n_e$) equal to 0.83[19S]. Thus, the radial variation of $\rho$ can then be directly inferred from the $n_e$ values, providing mass densities on the order of (2.1 – 5.6) x 10$^{-12}$ kg/m$^3$, which are, again, consistent with previous spectroscopically-derived measurements[20S].

**Focusing the analyses using a-priori knowledge:**
Throughout the main analyses documented in this Letter, an expanding series of circular annuli are employed to measure the azimuthally averaged radial variations of propagating magneto-acoustic wave phenomena as a function of distance from the umbral barycenter. This is implemented to ensure that the described techniques are readily applicable to the vast majority of solar active regions, irrespective of the shape of the underlying sunspot. However, this naturally begs the question of whether previous knowledge on coronal wave phenomena can help an astute researcher better constrain the errors associated with the derived magnetic fields, or whether a-priori knowledge is not required to derive accurate coronal field strengths. From examination of the $T_e$, *EM* and AIA 171Å intensity maps, it is clear that while the underlying sunspot may appear circularly symmetric, there are well-defined coronal structures manifesting as elongated fans in the north, west and south quadrants, with relatively cool characteristic temperatures (<1,000,000 K; Figure S2(c)). Furthermore, from examination of Movies S1 & S2 it becomes apparent that the most prevalent propagating wave phenomena occur along these well-defined coronal structures. Thus, a more focussed study would be to isolate those coronal features demonstrating the most pronounced compressive wave activity, and perform tailored analyses on these selected regions to characterize what, if any, improvements in the quantification of the coronal magnetic fields can be achieved.

Thus, to ensure that excessive noise is not incorporated into the analysis, it is desirable to isolate the coronal fan structures from background features. We therefore utilise a series of expanding arcs to more accurately confine plasma brightly emitting in the 171Å channel and to measure coronal variations more readily as a function of radial distance from the umbral barycenter. Similar to the expanding-annuli method used in the main text, each arc is 2 pixels wide (870 km), with subsequent arcs spaced further away by 1 pixel from the preceding measurement location. Some overlap between adjacent arcs is again chosen to provide continuous radial coverage of the measured parameters, while still maintaining high pixel numbers to improve statistics. A series of sample arcs are displayed in Figure S1(c), where it is clear that the prominent coronal fans are accurately encompassed by the series of expanding arcs.

The time–distance diagrams produced through either full annuli averaging or arc-based averaging are dominated by the propagating magneto-acoustic disturbances manifesting in the coronal fan structures. This is a direct consequence of the south-easterly sunspot quadrant displaying no prominent wave phenomena (Figure S1 and Movie S1), and as a result, these locations do not contribute (constructively or destructively) to the resulting time–distance diagrams. Thus, the measured gradients, which correspond to the wave tube speeds, $c_t$, remain unchanged to within a fraction of a km/s. Interestingly, a comparison between Figure S1 panels (b) and (d) reveals that employing more focussed expanding arcs, instead of generalized circularly symmetric annuli, does not result in significant changes to the measurements of propagating wave phenomena as a function of distance from the umbral barycenter. It is noticeable that the power spectra for azimuthally averaged annuli are marginally broader than those averaged across the selected coronal fans, which is likely a result of the smoothing of wave signatures caused by averaging over more spatial pixels. However, this does not detract from or modify the dominant periodicities found at each radial distance from the umbral barycenter, with the resulting spectral peaks located at similar positions in Figure S1(b&d). These findings suggest that the calculations of the velocity–distance variables (see the **Converting velocity–period to velocity–distance** section above) are accurate, irrespective of whether focussed arc-based averaging or more generalized circular annuli are employed.

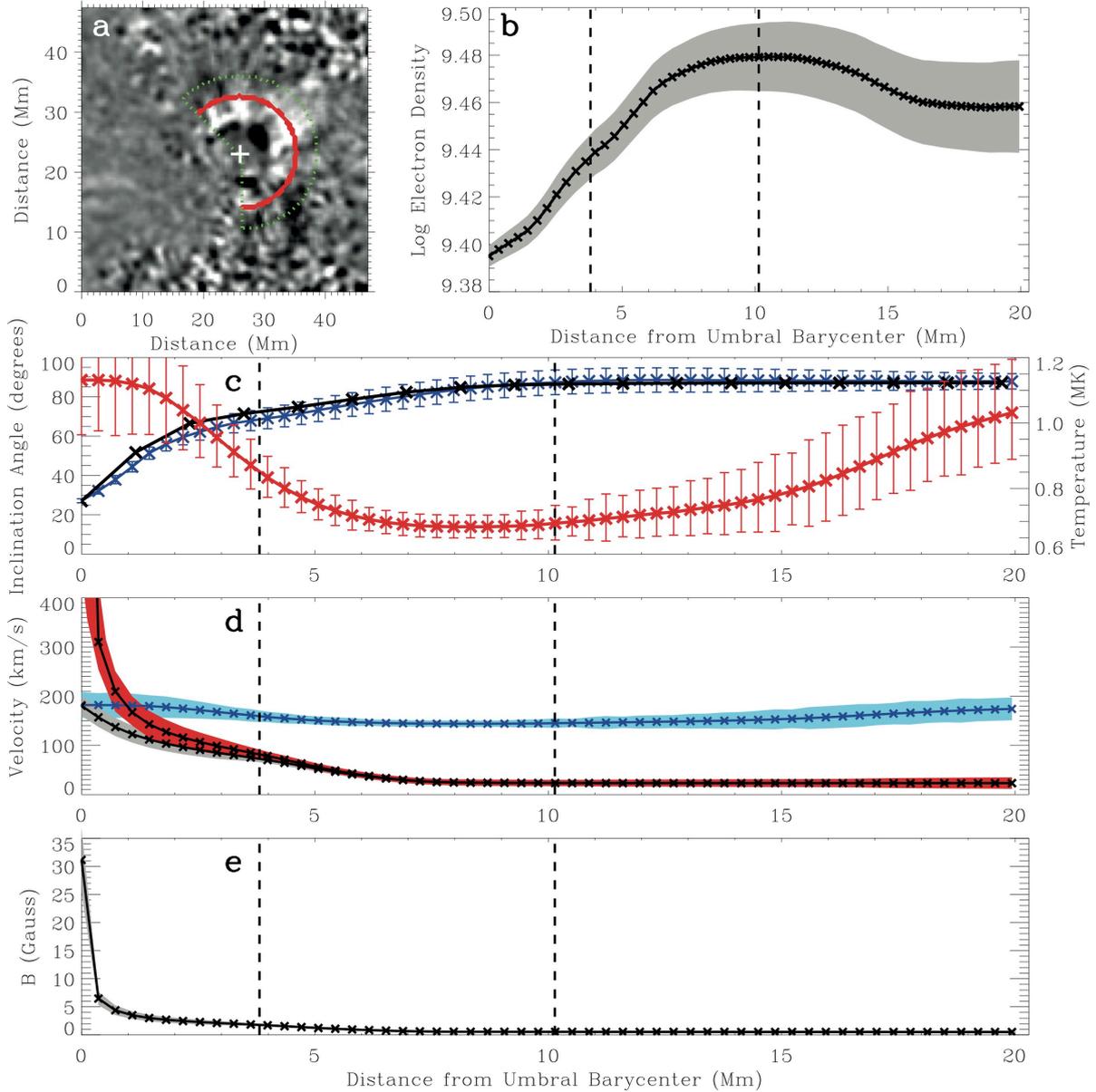

*Figure S2:* **Magnetic, temperature and velocity structuring of the sunspot plasma based on isolated coronal fan structures.**
*The panels displayed in this Figure are alternative representations of those already produced in the main Letter body, only now employing arc-based azimuthal averaging to focus the analyses based on a-priori knowledge of propagating magneto-acoustic waves in the corona. The following panel descriptions are repeated for completeness.* **(a)** *A snapshot of a 171Å intensity image having first been passed through a 4-minute temporal filter. The umbral barycenter is marked with a white cross, while the instantaneous peaks and troughs of propagating waves are revealed as white and black intensities, respectively, immediately surrounding the sunspot (see also Movie S2). The solid red line highlights an individual expanding arc, while the dotted green line outlines the spatial shape occupied by the series of the expanding arcs.* **(b)** *The arc-based azimuthally-averaged electron densities, $n_e$, displayed on a log-scale (units of $cm^{-3}$), where the vertical dashed lines represent the umbral and penumbral boundaries established from photospheric SDO/HMI continuum images at approximately 3.8 Mm and 10.1 Mm, respectively, from the center of the sunspot umbra.* **(c)** *Plots of the inclination angles of the coronal fan measured from STEREO–A images (black line) alongside the extrapolated magnetic field inclination angles (blue line), both transposed into the viewing perspective of the SDO spacecraft. The azimuthally-averaged plasma temperature (in MK) derived from DEM techniques is shown using a red line.* **(d)** *The measured tube speeds ($c_t$; black line with grey error contours) following compensation from inclination angle effects, the temperature-dependent sound speeds ($c_s$; blue line with light blue error contours) calculated from the local plasma temperature, and the subsequently derived Alfvén speeds ($v_A$; black line with red error contours).* **(e)** *The magnitude of the magnetic field strength, $|B|$, plotted as a function of radial distance from the umbral barycenter. Error estimates for all panels are computed in identical ways to those displayed in Figures 4&5.*

Important aspects to reconsider are the inclination angles of the magnetic fields in which the magneto-acoustic waves propagate along. Since the south-easterly quadrant of the sunspot does not display similarly structured coronal fans, it may introduce larger associated errors when averaging inclination angles over these differing atmospheric regions. This aspect becomes visible in Figure 4(c), where the error bars attached to the blue line reach a maximum of approximately ±10º, a consequence of averaging over a range of coronal structures. It can also be seen in Figure 4(c) that the separation between the azimuthally averaged extrapolated magnetic fields (blue line) and the physically measured inclination angles (using STEREO–A; black line) becomes visible when applying generalized circular annuli. Hence, can the focussed isolation of the coronal fan structures reduce the errors associated with the magnetic field inclination angles? Employing the same methodology detailed in the above **Compensating for inclination-angle effects in the $c_t$ measurements** section, only now utilizing arc-based averaging instead of full annuli averages, we find that the specific inclination angles as a function of radial distance from the umbral barycenter become slightly modified, but importantly, the associated error estimates also become reduced. Specifically, and most notably, the inclination angles remain unchanged within the confines of the sunspot umbra. It is well accepted that the magnetic field vectors inside sunspot umbrae are most vertical (i.e., dominated by components normal, $B_z$, to the solar surface), which remains valid whether azimuthally averaging within arc segments or circular annuli. Further away from the sunspot umbra, particularly within the penumbral regions, the magnetic field inclination angles with respect to the SDO spacecraft's line-of-sight increase at a faster rate (blue line in Figure S2(c)). This places the azimuthally averaged inclination angles much closer to the physically measured values using STEREO–A (black line), with a significant reduction in the associated errors – now on the order of ±5º. Again, as found for full annuli averages, at far distances from the umbral barycenter the arc-based segment averages unveil inclination angles that are close to 90º with respect to the SDO spacecraft's line-of-sight. Hence, focussing the analyses by selectively isolating specific coronal fan structures only modifies the magnetic field inclination angles by a small degree within the penumbral regions, but importantly allows for a reduction in the overall error estimates that propagate through the remaining magnetic field calculations.

Following the above results, the compensated tube speeds (grey values in Figure S2(d)) remain unchanged within the umbra and at far reaches from the umbral barycenter (i.e., beyond 10 Mm). Within the confines of the penumbra, the more rapid increase in the magnetic field inclination angles results in fractionally larger $\cos(\theta)$ terms, hence reducing the compensated tube speeds when compared to those calculated from full circular azimuthal averages (i.e., Figure 4(d)). However, this effect is negated by the fact that the isolated coronal fans are relatively cool (see, e.g., the red lines in Figures 4(c) and S2(c)), thus also producing a reduction in the associated sound speeds for these locations. As a result, the values of the deduced Alfvén speeds are almost identical to those obtained from full circular azimuthal averaging. By comparing the values documented in Figures 4(d) and S2(d), peak Alfvén speeds close to the umbral barycenter are found to be 1535 km/s and 1505 km/s following circular annuli and arc-based averaging, respectively, while the Alfvén speeds under both regimes reduce to approximately 25 km/s after a lateral distance of 7000 km.

The final step required before calculating the coronal magnetic field strengths is to estimate the local plasma density, ρ. From Figure 1 it is apparent that the cool coronal fan structures are also more dense than the plasma present in, e.g., the south-easterly quadrant of the sunspot. By computing the arc-based azimuthal average of the electron densities as a function of distance from the umbral barycenter (Figure S2(b)), it is clear that the resulting densities are indeed higher than those found using full circular annuli averages (Figure 5(b)). Both measurements are calculated using a value of 1300 km as the depth of the magneto-acoustic waveguides, which is consistent with a vast array of previously published work on magnetic coronal structures[15S,16S], in addition to our own independent measurements employing STEREO–A imaging data. We do not account for any expansion of the coronal fan structures with increasing lateral distance for three reasons. Firstly, the expansion is not readily apparent in our side-on STEREO–A images. While the STEREO–A spatial resolution (1.6 arcsec per pixel) is not as high as SDO/AIA observations (0.6 arcsec per pixel), a lack of visible expansion allows us to stipulate with some certainty that the expansion is minimal (at least less than one pixel), which corroborates previous statistical studies that document confined values over a multitude of spatial positions[15S,16S]. Secondly, the calculation of the magnetic field strength, $|B|$, depends on $\sqrt{\rho}$, which significantly reduces the impact of small-scale changes in the density, $\rho$. For example, if the coronal fan structure expands within the mid-penumbra by a single SDO/AIA pixel, then the depth will increase from 1300 km to approximately 1700 km, hence reducing the associated plasma density by a factor equal to 1.15. The consequence of this is a reduction in the estimated magnetic field on the order of 4%, which would decrease the previously estimated value from approximately 2.00 G in the mid-penumbra to around 1.92 G – something that would be lost in the already propagated error estimates. Thirdly, and finally, expansion is highly unlikely to create additional errors within the immediate confines of the sunspot umbra where the magnetic field strengths are calculated to be their largest values. Here, the magnetic waveguides will be at their narrowest and will likely remain within a single magnetic scale-height, which is estimated to be on the order of 25 Mm in the lower corona[21S]. Under these conditions the expansion of the coronal fan structures will be minimal, and as a result of the above relationships, the propagation of magnetic field strength errors as a consequence of waveguide expansion will also be insignificant.

Utilizing the arc-based azimuthal averages of the Alfvén speed and the plasma density, Figure S2(e) displays the resulting magnetic field estimates in the corona as a function of lateral distance away from the sunspot barycenter. Comparing to Figure 5(a), it is clear that the focused arc-based analyses produces marginally smaller error estimates on the magnitude of the magnetic field. This is a direct consequence of the reduced errors on the inclination angles (and hence compensated tube speeds), sound speeds (and resulting Alfvén speeds), in addition to the local plasma densities. Importantly, however, is the fact that the extreme ends of the magnetic field strengths remain unchanged, with values of approximately 32 G at the center of the sunspot umbra, decreasing to values on the order of 1 G after a lateral distance of 7000 km. Marginal variations are located towards the umbral/penumbral boundary where more pronounced inclination angle effects are present (see, e.g., Figures 4(c) and S2(c)). Ultimately, the remarkable agreement between generally applicable techniques (i.e., through circular azimuthal averaging) and those with a more focussed impetus (e.g., pre-selecting regions of interest using arc-based azimuthal averaging) reiterates the validity and reliability of the techniques documented in this Letter.

**Future directions:**
The observations presented here are employed to demonstrate a new technique to extract the spatial distribution of the coronal magnetic field based upon the omnipresent nature of propagating slow-mode waves in solar active regions. The AIA dataset provides an idealized testbed for our methodologies through the fact the sunspot is reasonably close to disk center where the vector magnetograms are most accurate, isolated from other strong magnetic polarities and therefore accurately represented by magnetic field extrapolations, by the pronounced coronal magnetic structures extending radially away from the underlying sunspot, and by the complementary on-limb imaging observations provided by STEREO–A. Future use of full stereoscopic observations (e.g., from the STEREO–A, STEREO–B and SDO spacecraft, or indeed the upcoming Solar Orbiter) would allow the inclination angles of the coronal fans to be established with a higher degree of precision.

This technique will also allow the magnitude of the magnetic fields to be determined in a complete two-dimensional spatial configuration. To facilitate this, the removal of annuli averaging would allow the measured tube speeds, $c_t$, to be calculated (see **Measuring the tube speeds from time–distance diagrams** above) along a well-defined one-dimensional distance slice through the dataset. If one end of the distance slice remained fixed at the spatial location of the umbral barycenter, with the opposite end rotated through 360 degrees, users of this technique would be able to accumulate a two-dimensional spatial configuration of the wave tube speeds. In addition to large-scale sunspot loop arcades and fan structures, these techniques can also be readily applied to singular and isolated coronal loops and plumes, providing they display measurable wave activity. Then, employing a pixel-by-pixel extraction of the local sound speeds, $c_s$, and plasma densities, $\rho$, would enable a full derivation of the two-dimensional Alfvén speeds, alongside the subsequent computation of the pixel-by-pixel magnitude of the coronal magnetic field strength, $B$. The magnetic field map displayed in Figure 5 portrays a homogeneous magnetic field as a result of the azimuthal averaging undertaken. Employing a pixel-by-pixel approach removes this condition and would allow small-scale inhomogeneities to be uncovered. Under this regime, the only constraint on the derived magnetic field is that associated with the 2-pixel resolution element of the selected instrument. For AIA observations, a 2-pixel spatial resolution corresponds to 1.2 arcsec (870 km), which would therefore be conducive for providing well-resolved information on the magnetic field strengths both along *and* across typical coronal loop structures, including singular and isolated loops.

The sunspot under investigation in this Letter displayed minimal activity and morphology during the full 75-minute extent of the observing sequence, and as a result, we employed time-averaged maps of the emission measure, *EM*, and temperature, $T_e$, when estimating local coronal plasma conditions. As the solar features remained invariably fixed within all AIA imaging channels throughout the duration of the dataset, the improved signal-to-noise that accompanies such temporal averaging was deemed beneficial under these conditions. Furthermore, in this regime Fourier-based techniques were employed to extract the propagating wave signatures from the EUV image sequences since they provide the highest possible frequency resolution. As highlighted in the main Letter text, a similar approach to measuring the magnitude of the coronal magnetic field strength may also prove reliable in the limit of rapidly evolving and highly dynamic solar active regions, where time resolution is of utmost importance. In order to facilitate this, time resolved *EM* and $T_e$ maps must be utilized. As the AIA instrument obtains all EUV images necessary for DEM calculations within a 12 s interval, changes in the local emission measures (i.e., affecting the plasma densities, temperatures and sound speeds) can be identified within similar time frames.

It may also be expected that the rapidly changing magnetic field configurations in, e.g., the buildup to a solar flare, may induce subtle fluctuations in the dominant wave periodicities and/or propagating tube speeds. Under this guise, the loss of temporal information when using more simplistic Fourier-based techniques may prove unsuitable for such dynamic studies. Fortunately, for these specific circumstances a wealth of alternatives to traditional Fourier techniques now commonly exist, including windowed Fourier, wavelet, Bayesian and Empirical Mode Decomposition analyses, many of which have been successfully applied to solar research topics in recent years[5,11,14,15,28,7S,10S-12S]. While basic Fourier techniques still retain the best frequency resolution, the other approaches incorporate the ability to employ time localized oscillatory functions continuous in both frequency and time to search for variable and/or transient oscillations. This opens up the possibility of being able to detect and reconstruct fluctuations in the propagating wave characteristics, in addition to the inherent changes in the local magnetic field strengths, on timescales as short as the lowest periodicity present in the data (approximately 1 minute for this sunspot). Of course, the application of such approaches still require the Nyquist criterion to be honored, whereby the minimum detectable oscillation is at twice the data sampling rate. For the purposes of AIA observations, images are obtained with a temporal cadence of 12 s, hence allowing oscillations with periods as short as 24 s to be conclusively detected. In the current observations, the shortest periodicities detected were on the order of 1 minute, and thus do not violate the Nyquist criterion. As a result, higher frequency fluctuations (i.e., near the immediate vicinity of sunspot umbrae) can be resolved very rapidly since approximately 5 samples will be recorded during a typical oscillation cycle. At the other end of the frequency spectrum where the periodicities are much longer (approximately 11 minutes at distances exceeding 15 Mm from the sunspot barycenter), any changes in the dominant periodicity will still begin to appear almost immediately due to the vast oversampling of the data in the frequency domain (i.e., an 11 minute oscillation contains 55 individual data points). A benefit of substantially oversampling the wave motions (both at low and high frequencies) is that it reduces aliasing in the reconstructed time series (e.g., for time-distance analysis), and hence allows oscillatory information to be extracted with a much decreased noise floor. As a result, time-resolved $T_e$ and $EM$ maps could be employed without annuli-based spatial averaging, alongside time-resolved $c_t$ measurements (e.g., using wavelet-based filtering), which would allow the two-dimensional mapping of the magnetic field to be accomplished with a cadence comparable to the highest frequency oscillations detectable (in this case 60 s).